\documentclass[twocolumn,amsmath,showpacs,amsfonts,aps,prc,floatfix]{revtex4}
\usepackage{graphicx}
\usepackage{bm}
\newcommand{\dn}{\frac{dN^\phi}{dy}} 
\newcommand{\pt}{<p_T^\phi>} 
\begin{document}

\title{Lattice based equation of state and transverse momentum spectra of identified particles in ideal and viscous hydrodynamics  }
\author{Victor Roy}
\email[E-mail:]{victor@veccal.ernet.in}
\affiliation{Variable Energy Cyclotron Centre, 1/AF, Bidhan Nagar, 
Kolkata 700~064, India}
\author{A. K. Chaudhuri}
\email[E-mail:]{akc@veccal.ernet.in}
\affiliation{Variable Energy Cyclotron Centre, 1/AF, Bidhan Nagar, 
Kolkata 700~064, India}

\begin{abstract}

Assuming that in Au+Au collisions, a baryon free fluid is produced,
 transverse momentum spectra of identified particles ($\pi$, $K$, $p$ and $\phi$),
 in evolution of ideal and viscous fluid is studied. Hydrodynamic evolution is governed by  a lattice based equation of state (EOS), where the confinement-deconfinement transition is a cross-over at $T_{co}$=196 MeV. Ideal or viscous fluid was initialised to reproduce $\phi$ meson multiplicity in 0-5\% Au+Au collisions. 
Ideal or minimally viscous ($\eta/s$=0.08) fluid evolution reasonably well explain the transverse momentum spectra of pion's, kaon's,  and $\phi$ meson's 
in central and mid-central Au+Au collisions. Description to the data is much poorer in viscous fluid evolution with $\eta/s \geq$0.12. 
 The model however under estimate proton production by  a factor $\sim$ 2.  
 \end{abstract}

\pacs{47.75.+f, 25.75.-q, 25.75.Ld} 

\date{\today}  

\maketitle

 

\section{Introduction}\label{sec1}
 
Relativistic hydrodynamics provides a convenient tool to analyse relativistic heavy ion collision data. It is assumed that a fireball is created in the collisions. Constituents of the fireball collide frequently to establish local thermal equilibrium sufficiently fast and after a certain time $\tau_i$, hydrodynamics become applicable. If the macroscopic properties of the fluid e.g. energy density, pressure, velocity etc. are known at the equilibration time $\tau_i$, the relativistic hydrodynamic equations can be solved to give the space-time evolution of the fireball till a given freeze-out condition such that interaction between the constituents is too weak to continue the evolution. 
Using suitable algorithm (e.g. Cooper-Frye) information at the freeze-out can be converted into particle spectra and can be directly compared with experimental data. Thus, hydrodynamics, in an indirect way, can characterize the initial condition of the medium produced in heavy ion collisions. Hydrodynamics equations are closed only with an equation of state (EOS). It is 
one of the most important inputs of a hydrodynamic model. Through this input macroscopic hydrodynamic models make contact with the microscopic world and one can investigate the possibility of phase transition in the medium. 
Most of the hydrodynamical calculations are performed with EOS with a 1st order phase transition. For example, in  \cite{QGP3}, 1st order EOS, EOS-Q was used to
successfully analyse 
a host of experimental data in Au+Au collisions at RHIC. In EOS-Q, the deconfined or the Quark-Gluon Plasma (QGP) phase is modeled by a bag equation of state of non-interacting 
quarks and gluons, the confined or the hadronic phase is modeled by a non-interacting gas of hadronic resonances.  
   Ideal hydrodynamics analysis with EOS-Q indicate that in central Au+Au collisions, at the equilibration time $\tau_i \approx$ 0.6 fm, 
  central energy density of the QGP fluid is $\varepsilon_i \approx$30 $GeV/fm^{-3}$ \cite{QGP3}. However, lattice simulations \cite{lattice,Cheng:2007jq} indicate that the confinement-deconfinement transition is neither a 1st nor a 2nd order phase transition, rather a cross-over at $T_{co}$=196 MeV.  
  It is then important that lattice based EOS are used in hydrodynamic analysis of RHIC data, more so when we are trying
to verify the lattice prediction about confinement-deconfinement transition. 
  
One also note that ideal hydrodynamic predictions for the  initial energy density or temperature of the fluid produced in Au+Au collisions is not creditable as dissipative effects are not included.
In hydrodynamics, initial energy density or temperature of the fluid is obtained by fitting experimental data on particle production, e.g. pion multiplicity, $p_T$ spectra etc., which essentially measure the final state entropy.
Unlike in ideal fluid evolution, where initial and final state entropy remains the same, in viscous fluid evolution entropy is generated. Consequently, to produce a fixed final state entropy, viscous fluid 
require less initial entropy density (or energy density) than an ideal fluid.

Recently, we have constructed an lattice based EOS and use it to explain the $\phi$ meson production in Au+Au collisions at RHIC \cite{Chaudhuri:2009uk}. 
Recent lattice simulation results \cite{Cheng:2007jq} were parameterised to obtain EOS of the deconfined state. The confined part of the EOS corresponds to that of a hadronic resonance gas with all the resonances with mass $m_{res} \leq $2.5 GeV. The deconfined and the confined part of the EOS were smoothly joined at a cross-over temperature $T_{co}$=196 MeV. In \cite{Chaudhuri:2009uk} it was shown that the lattice based EOS reasonably well explain the centrality dependence of $\phi$ mesons multiplicity ($dN^\phi/dy$), mean $p_T$ ($\langle p_T^\phi\rangle$) and integrated elliptic flow ($v_2^\phi$). From a simultaneous fit to  $dN^\phi/dy$, $\langle p_T^\phi\rangle$ and  $v_2^\phi$ an estimate of   the shear viscosity to entropy ratio was obtained, $\eta/s$=0.07 $\pm$ 0.03 $\pm$ 0.14, the first uncertainty is due to uncertainty in STAR measurements, the 2nd one is due to uncertain initial condition e.g. initial time varying between $\tau_i$=0.2-0.6 fm), freeze-out temperature varying between $T_F$=130-150 MeV, initial velocity ($v_r=tanh(\alpha r)$, $\alpha$=0.0-0.06), inaccuracy in hydrodynamic evolution code etc. 
  
 Strange meson $\phi$ constitute only a very small fraction of the total particles produced in Au+Au collisions. Particle production is dominated by pions,  kaons  and protons etc. For example, in central Au+Au collisions, pions   constitute nearly $\sim$80\% of the total particle yield, Kaons  $\sim$ 13\% and
protons $\sim$ 5\%. $\phi$ mesons contribute $\sim$ 2\% to the total yield.
It is then important to inquire whether or not hydrodynamics with the lattice based EOS is consistent with the experimental  data on other particles, e.g. pion, kaon, proton etc. The estimate of viscosity as obtained in   \cite{Chaudhuri:2009uk} will not be creditable unless the model also reproduces bulk of the particles, i.e. $\pi$, $K$, proton etc.   In the present paper, with the same parameters as in \cite{Chaudhuri:2009uk}, we have analysed transverse momentum spectra of identified particles, e.g. $\pi$, $K$, proton and $\phi$ in Au+Au collisions over a wide range (0-60\%) of collision centrality. In central and mid central collisions, hydrodynamic evolution of minimally viscous fluid best explain the data. Nearly equilvalent description is also obtained in ideal fluid evolution. However, description to the data in evolution with viscosity $\eta/s \geq 0.12$ is considerably poor than that in ideal or minimally viscous fluid.  
 
The paper is organised as follows: in section \ref{sec2}, we briefly describe the 
hydrodynamical equations used to compute the evolution of ideal and viscous fluid. 
 Construction of the lattice based equation of state is discussed in section \ref{sec3}. Simulation results   are discussed in section \ref{sec4}. Summary and conclusions are given in section \ref{sec5}.

\begin{figure}[t]
\vspace{0.3cm} 
\center
 \resizebox{0.35\textwidth}{!}{%
  \includegraphics{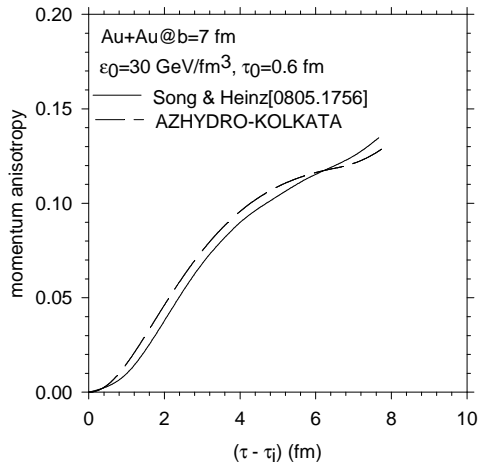}
}
\caption{
Viscous fluid ($\eta/s$=0.08) simulation for temporal evolution of momentum anisotropy in b=7 fm Au+Au collision at RHIC. The solid line is the simulation result from VISH2+1 \cite{Song:2008si} 
and the dashed line is the simulation result from AZHYDRO-KOLKATA.}\label{F1}
\end{figure}

\section{Hydrodynamical Equations, equation of state and initial conditions} \label{sec2}
\subsection{hydrodynamic equations}

 In Israel-Stewart's theory of 2nd order dissipative hydrodynamics, for a baryon free fluid, and neglecting bulk viscosity and heat conduction, the space-time evolution of a relativistic fluid is obtained by solving, 
 
\begin{figure}[t]
\vspace{0.3cm} 
\center
 \resizebox{0.35\textwidth}{!}{%
  \includegraphics{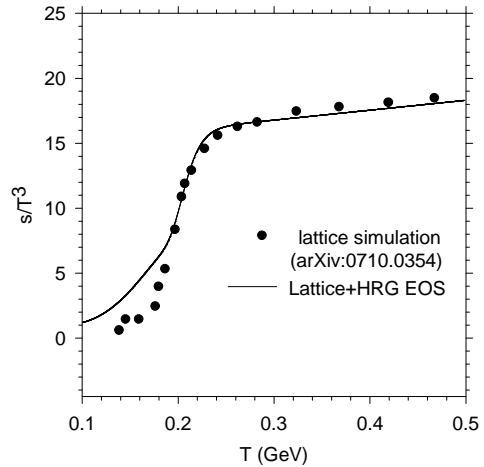}
}
\caption{Black circles are lattice simulation \cite{Cheng:2007jq} for entropy density. The black line is the parametric representation to the combination of lattice and hadron resonance gas EOS joined smoothly at cross over temperature $T_{co}$=196.0 MeV.
}\label{F2}
\end{figure}

\begin{eqnarray}  
\partial_\mu T^{\mu\nu} & = & 0,  \label{eq3} \\
D\pi^{\mu\nu} & = & -\frac{1}{\tau_\pi} (\pi^{\mu\nu}-2\eta \nabla^{<\mu} u^{\nu>}) \nonumber \\
&-&[u^\mu\pi^{\nu\lambda}+u^\nu\pi^{\nu\lambda}]Du_\lambda. \label{eq4}
\end{eqnarray}

Eq.\ref{eq3} is the conservation equation for the energy-momentum tensor, $T^{\mu\nu}=(\varepsilon+p)u^\mu u^\nu - pg^{\mu\nu}+\pi^{\mu\nu}$, 
$\varepsilon$, $p$ and $u$ being the energy density, pressure and fluid velocity respectively. $\pi^{\mu\nu}$ is the shear stress tensor. Eq.\ref{eq4} is the relaxation equation for the shear stress tensor $\pi^{\mu\nu}$.   
In Eq.\ref{eq4}, $D=u^\mu \partial_\mu$ is the convective time derivative, $\nabla^{<\mu} u^{\nu>}= \frac{1}{2}(\nabla^\mu u^\nu + \nabla^\nu u^\mu)-\frac{1}{3}  
(\partial . u) (g^{\mu\nu}-u^\mu u^\nu)$ is a symmetric traceless tensor. $\eta$ is the shear viscosity and $\tau_\pi$ is the relaxation time.  It may be mentioned that in a conformally symmetric fluid relaxation equation can contain additional terms  \cite{Song:2008si}.
Assuming boost-invariance, Eqs.\ref{eq3} and \ref{eq4}  are solved in $(\tau=\sqrt{t^2-z^2},x,y,\eta_s=\frac{1}{2}\ln\frac{t+z}{t-z})$ coordinates, with a code 
  "`AZHYDRO-KOLKATA"', developed at the Cyclotron Centre, Kolkata.
 Details of the code can be found in \cite{Chaudhuri:2008je,Chaudhuri:2008sj,Chaudhuri:2007qp,Chaudhuri:2008ed}. As shown in Fig.\ref{F1},
for similar initial conditions,  within 10\%, the code reproduces the temporal evolution 
 of   momentum anisotropy $\varepsilon_p=\frac{<T^{xx}-T^{yy}>}{<T^{xx}+T^{yy}>}$ of a QGP fluid as calculated by Song and Heinz \cite{Song:2008si}. 
 
\begin{figure}[t]
\vspace{0.0cm} 
\center
 \resizebox{0.45\textwidth}{!}{%
  \includegraphics{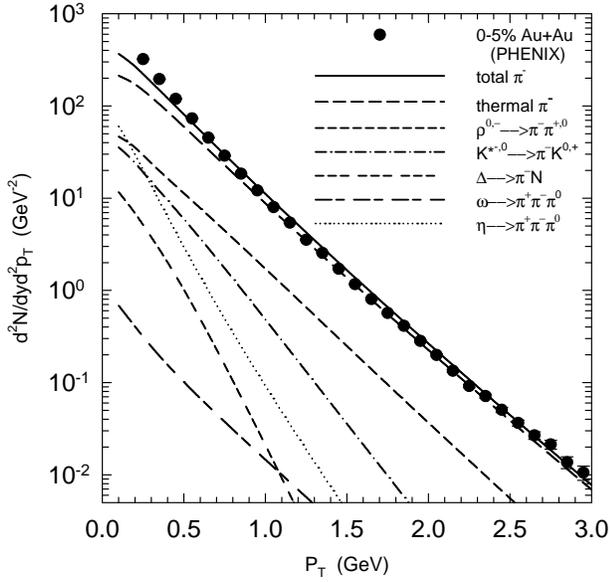}
}
\caption{Ideal hydrodynamic predictions for $p_T$ spectra of $\frac{\pi^+ +\pi^-}{2}$  in 0-5\% Au+Au collisions are compared with PHENIX data \cite{Adler:2003cb}. Thermal and decay pions are shown separately.}\label{F3A}
\end{figure}  
 
\subsection{Equation of state} \label{sec3}

Equation of state (EOS) is  one of the most important inputs of a hydrodynamic model. Through this input macroscopic hydrodynamic models make contact with the microscopic world.  
Most of the hydrodynamical calculations are performed with EOS with a 1st order phase transition.
Huovinen    \cite{Huovinen:2005gy} reported an 'ideal' hydrodynamic simulation with a cross-over transition. He concluded that the experimental data (e.g. elliptic flow of proton or antiproton) are better explained with EOS with 1st order phase transition than with EOS with 2nd order phase transition.
Huovinen    \cite{Huovinen:2005gy} used the 'thermal quasiparticle model' \cite{Schneider:2001nf}
to obtain EOS for the deconfined phase. For the confined phase he used the 
hadronic resonance gas model.

Recently, Cheng et al \cite{Cheng:2007jq} presented  high statistics lattice QCD results for the bulk thermodynamic observables, e.g. pressure, energy density, entropy density etc. The   simulations were performed with two light quarks and a heavy strange quark. The quarks masses are 'almost' physical, and corresponding pion mass is $m_\pi \sim$ 220 MeV. The strange quark mass was adjusted to physical value $m_K\sim$503 MeV.
In Fig.\ref{F2}, we have shown the simulation result for the entropy density \cite{Cheng:2007jq}. 
We have parameterise the entropy density as,  

\begin{equation}\label{eq1}
\frac{s}{T^3}=\alpha+[\beta+\gamma T][1+tanh\frac{T-T_c}{\Delta T}],
\end{equation}

In Fig.2,   the solid curve is a parameterisation   with $\alpha$=0.64, $\beta$=6.93,$\gamma$=0.55, $T_{c}$=196MeV,$\Delta$T=0.1$T_{c}$.
From the parametric form of the entropy density, pressure and energy density can be obtained using the thermodynamic relations,

\begin{figure}[t]
\vspace{0.0cm} 
\center
 \resizebox{0.45\textwidth}{!}{%
  \includegraphics{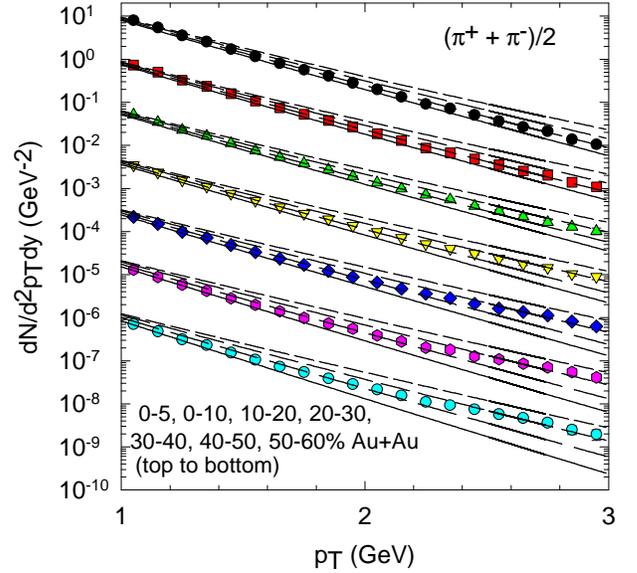}
}
\caption{(color online) PHENIX data \cite{Adler:2003cb} on the centrality dependence of 
$\frac{\pi^++\pi^-}{2}$   in 0-60\% Au+Au collisions.  
For better visibility experimental data beyond 0-5\% are divided by a factor of 10 respectively.
 The solid, dashed, medium dashed and long dashed   lines are  hydrodynamic predictions from ideal($\eta/s$=0.0) and viscous ($\eta/s$=0.08,0.12,0.16) fluid respectively.
   }\label{F3}
\end{figure}

\begin{eqnarray}  
  p(T)&=&\int_0^T s(T^\prime) dT^\prime \label{eq2a} \\
  \varepsilon(T)&=&Ts -p \label{eq2b}.
  \end{eqnarray}

We complement the lattice EOS \cite{Cheng:2007jq} by a hadronic resonance gas(HRG) EOS comprising all the resonances below the mass 2.5 GeV. The entropy density of complete EOS is obtained as

\begin{equation}  
  s=0.5[1+tanh(x)]s_{HRG}+0.5[1-tanh(x)]s_{LATTICE} \label{eq2c} \\
  \end{equation}

 \begin{table}[h]
\caption{\label{table1} Initial central energy density ($\varepsilon_i$) and temperature ($T_i$) of the fluid in b=0 Au+Au collisions, for different values of viscosity to entropy ratio ($\eta/s$). The predicted $\phi$ meson multiplicity
and mean $p_T$ are also noted. They should be compared with STAR measurements,
  ${\dn}_{ex}=7.95\pm0.74$ and ${\pt}_{ex}=0.977 \pm 0.064$.
} 
\begin{ruledtabular} 
  \begin{tabular}{|c|c|c|c|c|}\hline
$\eta/s$         & 0    & 0.08 & 0.12 & 0.16 \\  \hline
$\varepsilon_i (GeV/fm^3)$ & $35.5$ & $29.1$ & $25.6$ &  $20.8$ \\  
  & $\pm$ 5.0 & $\pm$ 3.6 & $\pm$ 4.0 &  $\pm$ 2.7  \\ \hline
$T_i$ (MeV) & 377.0 & 359.1 & 348.0 & 330.5     \\ 
  & $\pm 13.7$ & $\pm 11.5$ & $\pm 14.3$ & $\pm 11.3$   \\ \hline  
$\frac{dN^\phi}{dy}$ & 7.96 & 8.01 &  8.22 & 8.13\\ \hline
$<p_T\phi>$ & 1.019 & 1.062 &  1.111 & 1.174\\ \hline
 \end{tabular}\end{ruledtabular}  
\end{table}  

with x=$\frac{T-T_{c}}{\Delta T}$, $\Delta T=0.1T_c$. Compared to lattice simulation, entropy density in HRG drops slowly at low temperature, trace anomaly $\frac{\epsilon-3p}{T^{4}}$ drops faster in lattice simulation than in HRG model. It is difficult to resolve whether the discrepancy between lattice simulation at low temperature and HRG model is due to failure of HRG model at lower temperature or due to the difficulty in resolving low energy hadron spectrum on a rather coarse lattice \cite{Cheng:2007jq}.

\subsection{Initial conditions} \label{sec4}
  
Boost-invariant solution of Eqs.\ref{eq3} and \ref{eq4}  require initial conditions, e.g. transverse profile of the energy density ($\varepsilon(x,y)$) and
fluid four velocity ($u(x,y)$) and stress tensor $\pi^{\mu\nu}(x,y)$ at the initial time ($\tau_i$). Relaxation equation (Eq.\ref{eq4}) require to specify
the relaxation time $\tau_\pi$. A freeze-out prescription, e.g. freeze-out temperature $T_F$ is also needed.   In the present paper, we fix the initial
condition of the fluid as it was obtained in our analysis of $\phi$ mesons \cite{Chaudhuri:2009uk}.
At the initial time   $\tau_i$=0.6 fm, the fluid velocity is zero, $v_x(x,y)=v_y(x,y)=0$, the energy density of the fluid is distributed as,

\begin{equation}
\varepsilon({\bf b},x,y)=\varepsilon_i[0.75N_{part}({\bf b},x,y)+0.25 N_{coll}({\bf b},x,y)],
\end{equation}

\noindent where $N_{part}({\bf b},x,y)$ and $N_{coll}({\bf b},x,y)$ are transverse
profile of the participant and collision number distribution respectively, in an impact parameter {\bf b} Au+Au collision, calculated in a Glauber model. $\varepsilon_i$ is the central energy density in $b=0$ collisions.  The shear stress tensor is initialised to boost-invariant value.  
For the relaxation time we use Boltzmann estimate $\tau_\pi=3\eta/4p$.  
Freeze-out temperature is chosen to be $T_F$=150 MeV. 
The central energy density $\varepsilon_i$ is obtained by fitting 
$\phi$ multiplicity in 0-5\% Au+Au collisions \cite{Chaudhuri:2009uk}. The fitted values of central energy density and temperature are noted in table.\ref{table1}. As expected, the central energy density or temperature is reduced in more viscous fluid.

\begin{figure}[t]
\vspace{0.0cm} 
\center
 \resizebox{0.45\textwidth}{!}{%
  \includegraphics{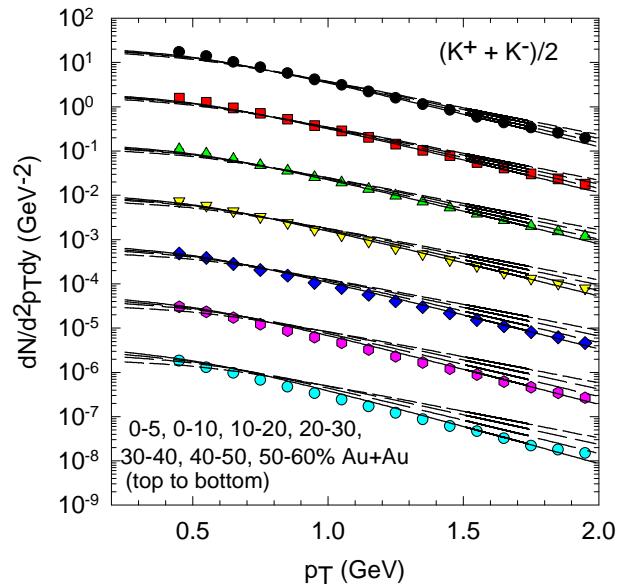}
}
\caption{(color online) same as in Fig.\ref{F3} but for kaons.   }\label{F4}
\end{figure}

\begin{table}[h]
\caption{\label{table2} The $\chi^2/N$ values, in ideal and viscous ($\frac{\eta}{s}$=0.08-0.16) evolution  for the $\pi$, $K$, $p$ and $\phi$ in different centrality ranges of Au+Au collisions are shown. In the last two rows,
$\chi^2/N$ values for the combined data sets, ($\pi$+$K$+$p$+$\phi$) 
and ($\pi$+ $K$+   $\phi$) in the centrality range 0-40\% are shown.} 
\begin{ruledtabular} 
  \begin{tabular}{|c|c|c|c|c|c|}\hline
particle & collision &  \multicolumn{4}{c|} {$\chi^2/N$}    \\ \cline{3-6}
species & centrality &$\frac{\eta}{s}$=0 &$\frac{\eta}{s}$=0.08 &$\frac{\eta}{s}$=0.12 & $\frac{\eta}{s}$=0.16\\ \hline
& 0-5\%&     39.18&      9.29&     67.14&    273.04\\
&0-10\%&     39.58&      7.38&     60.41&    262.07\\
&10-20\%&     47.63&      5.29&     47.73&    240.09\\
$\frac{\pi^++\pi^-}{2}$&20-30\%&     70.58&      7.38&     79.20&   385.80\\
&30-40\%&    51.49&       12.80&   135.38&   520.60\\
&40-50\%&    42.47&     23.63&    187.18&    595.89\\
&50-60\%&     38.18&     25.70&    186.38&    518.10\\ \hline
&0-5\%&      8.62&      4.43&     37.80&     99.31\\
&0-10\%&      7.00&      5.23&     43.11&    112.11\\
&10-20\%&      2.89&     12.83&     74.33&    176.19\\
$\frac{K^++K^-}{2}$&20-30\%&      6.00&     41.10&    175.10&    382.11\\
&30-40\%&     14.35&     89.61&    292.21&    589.08\\
&40-50\%&     27.92&    136.01&    391.16&    725.67\\
&50-60\%&      27.79&    137.53&    388.47&    675.92\\ \hline
&0-5\%&    120.91&     96.16&     72.55&     63.78\\
&0-10\%&    114.09&     89.92&     67.06&     59.03\\
&10-20\%&     90.15&     67.05&     46.84&     42.55\\
$\frac{p+\overline{p}}{2}$&20-30\%&     96.19&     62.74&     37.76&     40.67\\
&30-40\%&     61.57&     33.39&     17.71&     32.55\\
&40-50\%&     35.03&     13.19&      8.70&     35.75\\
&50-60\%&      19.74&      4.12&      6.03&     32.25\\ \hline
&0-5\%&     20.27&     31.49&     61.83&    102.17\\
&0-10\%&      8.94&     12.15&     22.76&     38.42\\
&10-20\%&      4.01&      5.51&     19.01&     45.91\\
$\phi$&20-30\%&     11.33&     13.94&     29.95&     60.46\\
&30-40\%&     22.25&     33.01&     64.11&    114.77\\
&40-50\%&     40.38&     71.75&    141.06&    246.70\\
&50-60\%&      49.00&     90.62&    169.84&    261.76\\ \hline
($\pi+K+p+\phi$)& 0-40\%& 41.85& 32.03&72.60& 182.03\\
($\pi+K+\phi$)& 0-40\%& 23.61& 19.43&80.67& 226.80\\
\end{tabular}\end{ruledtabular}  
\end{table} 

\section{Results}

With the initial conditions   as described above, we have computed transverse momentum spectra of pions, kaons, protons and $\phi$ mesons. In Fig.\ref{F3A}, predicted pion spectra from ideal fluid evolution in 0-5\% Au+Au collisions are shown. Thermal pion's and decay pions are shown separately.   Decay pions contribute mainly at low $p_T <1 GeV$. Decay and thermal pions  together (the solid line) reasonbly well explain the PHENIX data \cite{Adler:2003cb}. Note that initial condition of the fluid was not tuned for $\pi$ mesons, Rather it was tuned to fit $\phi$ meson multiplicity in 0-5\% collisions.
 
In Fig.\ref{F3}, present model predictions for $\frac{\pi^+ +\pi^-}{2}$,   in 0-60\% Au+Au collisions are compared against the PHENIX data \cite{Adler:2003cb}. 
For the ease of computation and with the understanding that resonances contribute mainly at low $p_T <$ 1 GeV, we have omitted decay pions. Accordingly spectra are shown only in the $p_T$  range $1\leq p_T \leq 3$ GeV.  In Fig.\ref{F3}, the colored symbols are the PHENIX data \cite{Adler:2003cb}.  
The solid, dashed, medium dashed and long dashed lines are hydrodynamics model predictions for the spectra in ideal fluid and in viscous fluid with viscosity to entropy ratio $\eta/s$=0.08, 0.12 and 0.16 respectively. At large momentum pion yield increases with viscosity.  
From Fig.\ref{F3}, it is evident that in central and mid-central collisions, both the ideal and minimally viscous fluid reasonably well explains the data. Data are over predicted in fluid evolution with $\eta/s$=0.12 and 0.16.  To obtain a quantative idea of fit to the data in ideal and viscous hydrodynamics,   we have computed $\chi^2/N$.
 They are noted in table \ref{table2}. For all the collisions centrality, minimum $\chi^2/N$ is obtained in fluid evolution with the ADS/CFT limit of viscosity $\eta/s$=0.08. $\chi^2/N$ is comparatively large in ideal fluid. Compared to ADS/CFT limit of viscosity, fit to data is considerably poor evolution with viscosity, $\eta/s$=0.12 and 0.16. We also note that irrespective of viscosity, quality of fit gets poorer as the collisions become more and  more peripheral.
 For example, in minimally viscous fluid, upto collision centrality 30-40\%, $\chi^2/N\sim$ 10, but beyond 30-40\% collision centrality, $\chi^2/N$ increases by a factor of $\sim$2. 30-40\% Au+Au collisions corresponds to average impact parameter b=8.3 fm. It appear that in a hydrodynamic model, pion spectra is explained only in  $b\leq $8.3 fm Au+Au collisions.  

\begin{figure}[t]
\vspace{0.0cm} 
\center
 \resizebox{0.45\textwidth}{!}{%
  \includegraphics{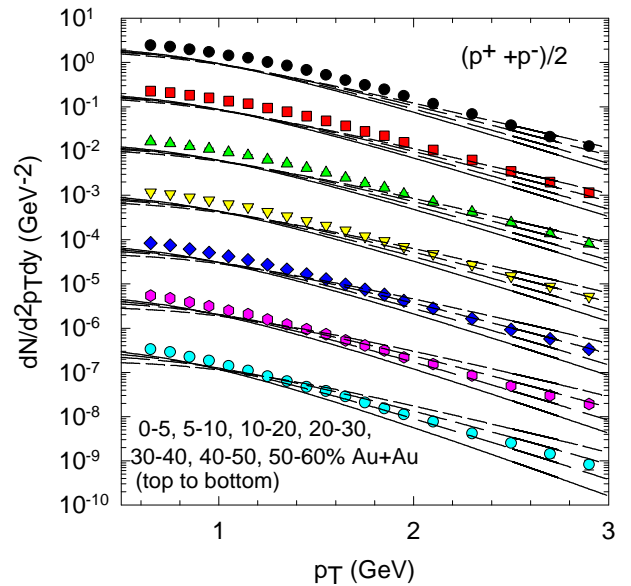}
}
\caption{(color online) same as in Fig.\ref{F3} but for protons. 
   }\label{F5}
\end{figure}

 Compared to pions, resonance production of kaon is much less.
In Fig.\ref{F4}, PHENIX data  \cite{Adler:2003cb} for $\frac{K^+ +K^-}{2}$ are
compared with hydrodynamic model predictions.  Fit to the data in the entire $p_T$ range are shown. Kaon spectra in the centrality range 0-40\% are also well explained in ideal and minimally viscous fluid evolution. As it was for pions, in more viscous fluid ($\eta/s \geq$0.12) description deteriorates. Again, to obtain a quantative idea of fit to the data, we have computed $\chi^2/N$. To be consistent, with the analysis of pion spectra, in the $\chi^2$ analysis, data only in the $p_T$ range $1 \leq p_T \leq 3$ GeV are included. Values  are  noted in table.\ref{table1}. Unlike for pions, best fit to the kaon data is obtained in ideal hydrodynamics. Viscous hydrodynamic gave comparatively poorer fit.

In Fig.\ref{F5},   hydrodynamic model prediction for proton spectra are compared with the PHENIX data  \cite{Adler:2003cb} for $\frac{p + {\overline p}}{2}$. Interestingly, both ideal or viscous hydrodynamics, with the lattice based EOS
fails to reproduce the proton data. In central or mid-central collisions, data are underpredicted by a factor of $\sim$2. Proton spectra in peripheral collisions are comparatively better fitted.     It is reflected in the $\chi^2/N$ analysis also (see table.\ref{table2}). Minimum $\chi^2/N$   for the proton spectra is order of magnitude larger than that obtained for pion or kaon spectra.
We have neglected resonance production. However, resonance contribution to  proton is very small and   even if resonances are included, model predictions will not agree with the experiment.
Inability of the model to correctly predict proton spectra is possibly due to the neglect of baryons in the model. We have assumed a baryon free fluid. However, the matter produced in Au+Au collisions at RHIC are not entirely baryon free. Note that with baryons in the fluid, proton production will depend exponentially on the chemical potential. Even a small chemical potential at the freeze-out will considerably enhance proton production. Inability of the model to correctly predict $p_T$-spectra of proton may also be due to the equation of state used presently.  
The equation of state   is a combination of  a recent lattice simulation \cite{Cheng:2007jq} and non interacting hadron resonance gas.  The lattice simulation was performed with two light quark and a heavier strange quark. Light quark masses that are nearly twice the physical masses, the strange quark mass was adjusted to its physical value. Consequently, pion is heavy,  $m_\pi\approx$220 MeV and kaon mass is physical $M_K \approx$503 MeV \cite{Cheng:2007jq}. Though it is not mentioned in \cite{Cheng:2007jq}, we expects that protons are also heavier than the  physical ones. Then in the lattice based EOS  the confined phase is a fluid of heavy pions and protons but physical kaons. At the freeze-out fluid cells will contain less number of heavier protons than it would have otherwise and proton production will be reduced.
Pion production will also be reduced, however, being a lighter particle, the effect of heavy pion mass will be less pronounced. For example, if we approximate $dN/dyd^2p_T \propto exp(-\sqrt{m^2+p^2_T}/T)$,  then in the $p_T$ range 1-3 GeV, for $\sim$ 50\% increase in pion mass,  production is reduced only by 3-10\%. For a similar increase in proton mass production is reduced by 60-90\%. Indeed, even if proton is only 20\% heavier than physical proton, proton production is reduced by 30-60\% in the $p_T$ range 1-3 GeV. More detailed study is needed to understand the proton spectra.
 
 \begin{figure}[t]
\center
 \resizebox{0.45\textwidth}{!}{%
  \includegraphics{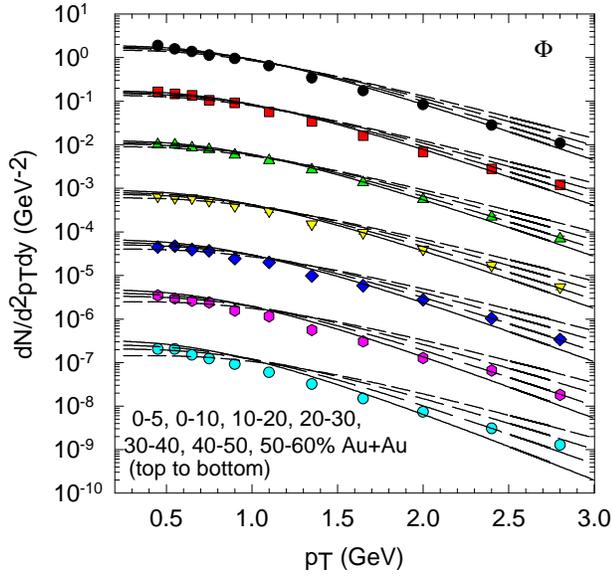}
}
\caption{(color online) same as in Fig.\ref{F3} but for $\phi$ mesons.
    }\label{F6}
\end{figure} 

As mentioned earlier, we have initialised the fluid (ideal or viscous) to reproduce $\phi$ meson multiplicity in 0-5\% Au+Au collisions.  In Fig.\ref{F6},
$\phi$ meson's $p_T$ spectra are studied. As before, the solid, dashed, medium dashed and short dashed lines are from evolution of fluid with $\eta/s$=0,0.08, 0.12 and 0.16 respectively. The filled circles are from the STAR experiment  \cite{Abelev:2007rw}. For the $\phi$ data also, 
$p_T$ spectra are best explained in   ideal fluid evolution. Comparatively poor description is obtained in viscous evolution. It is evident also from the $\chi^2$ values in table.\ref{table2}. In all the collision centrality, $\chi^2/N$ is minimum in ideal fluid evolution. $\chi^2/N$'s are comparatively larger in minimamly viscous fluid evolution. For  $\eta/s$=0.12 or 0.16, compared to ideal fluid, in viscous fluid evolution, $\chi^2/N$ increases by a factor of 3-6.

\begin{figure}[t]
\center
 \resizebox{0.45\textwidth}{!}{%
  \includegraphics{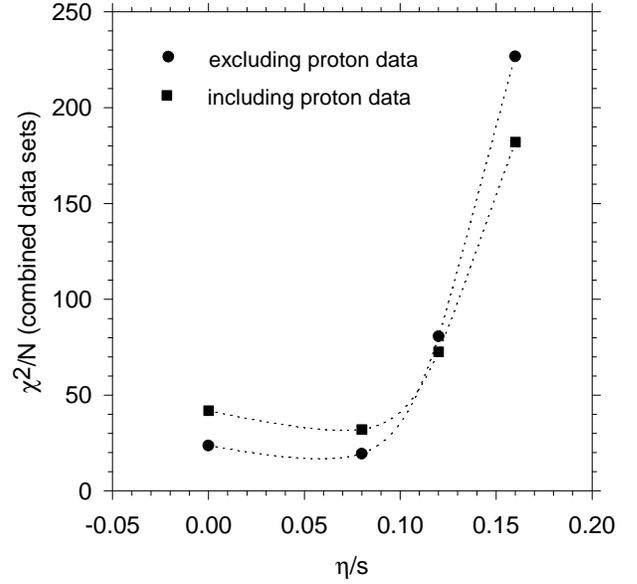}
}
\caption{Filled squares are the $\chi^2/N$ for the combined data sets   ($\pi$, $K$, $\phi$ and $p$) as a function of viscosity to entropy ratio ($\eta/s$).
The filled circles are the same when proton data are excluded. Experimental data only upto 30-40\% collision centrality are included.
    }\label{F7}
\end{figure}

In Fig.\ref{F7}, we have shown the $\chi^2/N$ values for the combined data sets;
$\pi$, $K$, $p$ and $\phi$. Collision centrality upto 30-40\% are included only. As indicated earlier, hydrodynamic description to the data gets poorer beyond this
collision centrality. In Fig.\ref{F7}, the filled square are the $\chi^2/N$ of the combined data sets, as a function of viscosity $\eta/s$. $\chi^2/N$ analysis definitely indicate that identified particle $p_T$ spectra do not demand large viscosity. Minimum $\chi^2/N\approx 32$ is obtained in minimally viscous evolution. But comparable description to the data is also obtained in ideal fluid evolution.
In Fig.\ref{F7}, the filled circles are the $\chi^2/N$ when proton data are excluded from the analysis. As noted earlier,  proton spectra are not well reproduced in the model. If proton data  are excluded, $\chi^2/N$ values improves. Again the best fit to the combined $\pi$, $K$ and $\phi$ data is obtained in minimally viscous fluid evolution, $\chi^2/N\approx 19$. Ideal hydrodynamics give comparable fit. The results are consistent with recent estimate of QGP viscosity \cite{Chaudhuri:2009uk}. In \cite{Chaudhuri:2009uk} analysing $\phi$ meson data, it was concluded that nearly perfect fluid is produced in Au+Au collisions at RHIC energy. Transverse momentum spectra of identified particles also lead to similar conclusions, in Au+Au collisions, a nearly perfect fluid is produced.
 
 Before we summarise our results, it is important to mention that we have neglected bulk viscosity. Experimental data, which include the effect of bulk viscosity, if there is any.  In general, bulk viscosity is an order of magnitude smaller than shear viscosity. But in QGP, it is possible that near the cross-over temperature,
bulk viscosity is large  \cite{Kharzeev:2007wb,Karsch:2007jc}. Effect of bulk viscosity on particle spectra and elliptic flow is studied in \cite{Monnai:2009ad}. It appears that even if small, bulk viscosity can have visible effect on particle spectra and elliptic flow. Neglect of bulk viscosity, will artificially  increase the effect of (shear) viscosity. In other word, if bulk viscosity is included, comparable fits to the data can be obtained with still lower value of $\eta/s$.  
 
\section{Summary and conclusions} \label{sec5}

To summarise, in a hydrodynamical model, where the evolution is governed by a lattice based equation of state with a confinement-deconfinement cross-over at temperature $T_{co}$=196 MeV, we have analysed the   transverse momentum spectra of identified particles, e.g. pions, kaons, protons and $\phi$ mesons. It is assumed that Au+Au collisions produce a   'baryon free'  ideal/viscous fluid.
Ideal or viscous ($\eta/s$=0.08-0.16)   fluid was initialised 
  to reproduce $\phi$ meson multiplicity in a central (0-5\%) Au+Au collision.
Hydrodynamic evolution of the ideal or minimally viscous ($\eta/s$=0.08)  fluid, initialised to reproduce $\phi$ multiplicity in 0-5\% Au+Au collisions,  reasonably well reproduces transverse momentum spectra of $\pi$, $K$ and $\phi$ in central and mid-central collisions. In peripheral collisions, 40-50\% and beyond, the description to the data gets poorer. Description to the that data is also poor in evolution of fluid with viscosity larger than the ADS/CFT limit.   Hydrodynamical evolution of baryon free ideal or viscous fluid  however do not generate enough protons to agree with experiment. Proton spectra are underpredicted by a factor of 2. Poor fit to the proton data      is possibly due to the neglect of baryons in the model. Fluid produced in Au+Au collisions at $\sqrt{s}$=200 GeV is not entirely baryon free. It is expected that the fits to proton data will improve if baryons are included in the model. Poor fit to proton data  may also be due to comparatively large light quark masses in lattice simulation. Light quarks are approximately twice the mass of physical quarks, consequently protons are heavy. More detailed study is need to sort out the issue. 
Our analysis also indicate that the transverse momentum spectra of the combined data set, ($\pi$, $K$, $p$ and $\phi$) or  ($\pi$, $K$, and $\phi$), in 0-40\% collision centrality are best explained in minimally viscous fluid. Nearly equilvalent description is obtained in ideal fluid evolution. Data definitely reject large viscous fluid, $\eta/s \geq$0.12.

\end{document}